\documentclass[referee,pdflatex,sn-nature]{sn-jnl}

\usepackage{siunitx}
\usepackage[normalem]{ulem} 
\usepackage{comment}
\usepackage{gensymb} 
\usepackage{graphicx}%
\usepackage{multirow}%
\usepackage{amsmath,amssymb,amsfonts}%
\usepackage{amsthm}%
\usepackage{mathrsfs}%
\usepackage[title]{appendix}%
\usepackage[dvipsnames]{xcolor}%
\usepackage{textcomp}%
\usepackage{manyfoot}%
\usepackage{booktabs}%
\usepackage{algorithm}%
\usepackage{algorithmicx}%
\usepackage{algpseudocode}%
\usepackage{listings}%
\usepackage{geometry}
\usepackage{pdfpages}
\usepackage{hyperref}
\hypersetup{
    colorlinks=true,
    linkcolor=black,
    filecolor=black,      
    urlcolor=black,
    citecolor=black,
    }
\begin{document}

\title{Chains of nanoparticles for flat-band emission and lasing}

\author[1]{Rebecca Heilmann}

\author[1]{Joel Lehikoinen}

\author[1]{Sioneh Eyvazi}

\author[1]{Evgeny A. Mamonov}

\author*[1]{\fnm{Päivi} \sur{Törmä}}\email{\textcolor{black}{paivi.torma@aalto.fi}}

\affil[1]{\orgdiv{Department of Applied Physics}, \orgname{ Aalto University School of Science}, \orgaddress{\street{P.O. Box 15100}, \postcode{Aalto FI-00076}, \country{Finland}}}

\date{\today}
\abstract{
Controlling light–matter interactions is central to photonic technologies ranging from lasers to optical information processing. Suitably designed photonic structures give rise to flat bands, where the density of states diverges, and group velocity goes to zero, allowing light localization. These properties make flat bands attractive for lasing; however, designing photonic structures supporting flat bands is challenging. Here, we introduce long-ranged coupled chains of nanoparticles that support totally flat bands extending over the full angular range. We demonstrate flat-band lasing in single chains and explain the transition to $\Gamma$-point lasing as the number of chains is increased. Moreover, we show partially coherent emission from square and triangular two-dimensional chain lattices. The excited modes depend on the pump power and polarization. Our results establish chain lattices as a versatile platform for exploring flat band-lasing and suggest new routes toward narrowband, linearly polarized, and bright light sources with tailored coherence.
}
\maketitle

Periodic photonic structures provide a means to tailor light--matter interactions by engineering the underlying band structure. While most photonic bands are dispersive, suitably designed lattices can host flat bands, i.e., bands of dispersionless modes whose energy remains constant over a range of momenta. Photonic and exciton--polariton systems with flat bands have attracted interest~\cite{leykam2018artificial, leykam2018perspective, Danieli2024} because they enable experimental realizations of Hamiltonians associated with exotic topological many-body phenomena, including fractional quantum Hall effect~\cite{Bergholtz2013}, superconductivity~\cite{Torma2022}, and ferromagnetism~\cite{Lieb1989}. Their nondispersive nature also gives rise to a diverging density of states, vanishing group velocity, and compact localized eigenstates arising from destructive interference~\cite{leykam2018artificial}, which in turn strengthen nonlinear optical processes~\cite{Ning2023, Sun2025}, give rise to slow-light effects~\cite{Baba2008}, and offer control over and enhancement of emission~\cite{munley2023visible} and absorption~\cite{choi2024nonlocal}. 

The above-described properties make flat band platforms particularly attractive for coherent light generation: localized states act as intrinsic optical cavities, low group velocity increases the photon lifetime and effective quality factor ($Q$ factor), and the high density of states increases emission rates, reducing the lasing threshold. However, designing flat band photonic systems is challenging, as long-range couplings, typical in photonic systems, tend to induce dispersion. Coherent emission from a flat band was accordingly first observed in exciton--polariton systems with Lieb or kagome tight-binding (nearest-neighbor coupled) lattices~\cite{baboux2016bosonic, klembt2017polariton, whittaker2018exciton, harder2020exciton, harder2021kagome, scafirimuto2021tunable}. For photonic systems, a theoretical description of a flat-band laser in an evanescently coupled (tight-binding) system was proposed by Longhi~\cite{longhi2019photonic}. By combining tight-binding and long-range effects, multiple scattering theory has been used to predict that 1D chains of high-index spheres with a specific chain spacing give rise to a flat band~\cite{Hoang2024}. Another noticeable system showing lasing is a moiré lattice realized by two twisted hexagonal geometries~\cite{Mao2021}, where twist-induced destructive interference creates a high-$Q$ flat band nanocavity. This platform was subsequently extended to reconfigurable nanolaser arrays with programmable emission patterns~\cite{Luan2023}. Other approaches have sought to leverage the inherently high $Q$ factors of quasi-bound states in the continuum (for brevity, BICs) in conjunction with flat bands. Eyvazi et al.~employed a silicon metasurface to couple out guided modes, whose dispersion is flattened by the contrast between the effective refractive index of the guided mode and its surrounding~\cite{Eyvazi2025}. They also observed symmetry-protected and accidental BICs on the flat band. Do et al.~carefully tuned the interaction of four guided modes via the anisotropy of a rectangular metasurface to locally flatten a band with a BIC at the $\Gamma$ point, leading to a flat mode with a $Q$ factor over 9000 in the visible range~\cite{Do2025}. Cui et al. designed a lattice featuring a symmetry-protected BIC at the $\Gamma$-point surrounded by accidental BICs yielding a flat band in THz spectral region~\cite{Cui2025}.

In the companion manuscript~\cite{Lehikoinen2026}, we show by theory that one-dimensional (1D) chains of nanoparticles and two-dimensional (2D) lattices constructed from them, referred to as chain lattices, host flat bands. Contrary to the previous photonic flat-band realizations~\cite{Mao2021, Luan2023, Eyvazi2025, Do2025, Cui2025}, these flat bands arise purely from diffraction in a non-tight-binding (long-range coupled) system. Furthermore, they extend over all momenta. Here, we investigate the optical properties of selected chain lattice geometries. We present how chain lattices transition from a flat band hosting structure to a regular 2D square lattice with TE- and TM-polarized surface lattice resonances (SLRs) as the array size changes in terms of dispersion and lasing. We study emission from square and triangular 2D chain lattices, and show how the spatial- and momentum-space emission patterns depend on the array geometry, pump intensity, and its polarization. We find coherence within individual chains, while the 2D lattice as a whole produces bright incoherent radiation. Our findings establish single chains and their lattices as a versatile platform for realizing both coherent and incoherent emission from flat bands.

We realized the lattices under study using metallic plasmonic nanoparticles embedded in a gain material. Plasmonic nanoparticle arrays~\cite{Kravets2018,wang2018rich} host collective modes called surface lattice resonances, which combine diffractive orders and single nanoparticle resonances, and are well suited for realizing the essential features of the chain lattice flat bands proposed in~\cite{Lehikoinen2026}. Moreover, the SLRs can provide optical feedback for the gain medium, resulting in lasing~\cite{Suh2012, Zhou2013, hakala2017lasing,freire2025plasmonic}. The extinction and emission spectra of the arrays are studied using momentum-resolved spectroscopy, and the spatial coherence of the emission is investigated via Michelson interferometry. 



\begin{figure*}
    \centering
    \includegraphics[width=\textwidth]{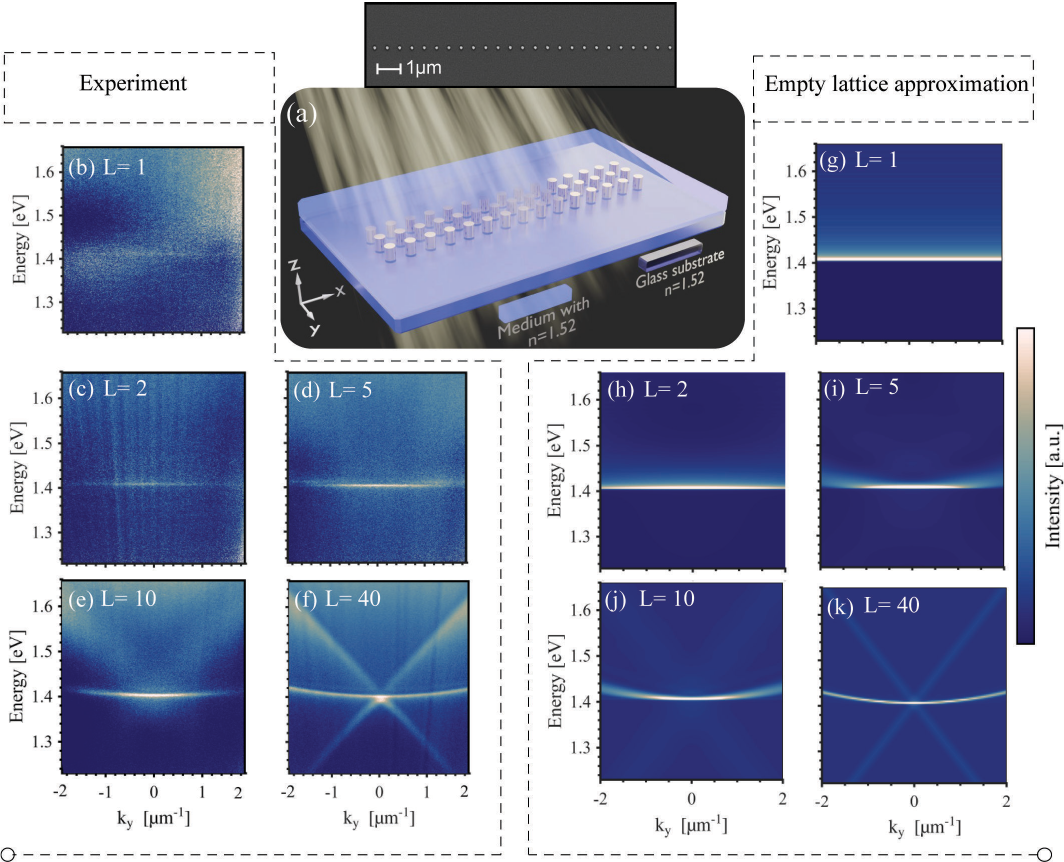}
    \caption{Experimental measurement of the extinction. (a) Scheme of a number ($L$) of parallel chains. The case $L=3$ is depicted in the schematic, while the inset shows a scanning electron microscope image of a single chain ($L=1$). Measured extinction [panels (b)--(f)] and band structure calculated using the empty lattice approximation [panels (g)--(k)] of $L = \text{1, 2, 5, 10, and 40}$ lines of square lattice with a periodicity of 580~nm. Particles were gold nanocylinders with a diameter of 120~nm and height of 50~nm in an index-symmetric background with $n = 1.52$, and no polarization filters were used in the measurement. The first Brillouin zone (for an $L \rightarrow \infty$ system) extends between $k_y \approx \pm \SI{5.4}{\per\micro\metre}$.}
    \label{f:dispersion_chains}
\end{figure*}

We first consider the band structures of arrays that are effectively infinite in the $x$ direction with the periodicity of $p$ and have $L$ chains (rows) in the $y$ direction with the same distance $p$ between them, see Fig.~\ref{f:dispersion_chains}(a). Intuitively, for a single chain, we expect its dispersion $E(k_y)$ at $k_x = 0$ to be flat, because the periodic structure in the $x$ direction permits constructive interference only at specific energies and wavevectors $\mathbf{k}_x$, whereas the lack of structure in the $y$ direction forces the band to be flat in that direction. On the other hand, in the limit of large $L$, we expect the dispersion to approach that of a square lattice, which does not have any flat bands, instead it shows dispersive TE- and TM-polarized SLR modes. This raises the question at which values of $L$ does the band structure transition from the flat dispersion of a single chain to that of the square lattice? We answer this question by studying arrays of gold nanoparticles using momentum-resolved transmission spectroscopy and by calculating the band structures of the lattices using the empty lattice approximation (for more information on sample fabrication, characterization, and theory, see Supplementary Information).

Figure~\ref{f:dispersion_chains} shows the experimentally obtained dispersions [panels (b)--(f)] and band structures calculated using the empty lattice approximation [panels (g)--(k)] of lattices with $L = 1, 2, 5, 10, \text{ and } 40$ chains, respectively. The chains consist of cylindrical gold nanoparticles with a diameter of 120~nm and a height of 50~nm with a chain periodicity of 580~nm. The periodicity was chosen such that the energy of the flat band overlaps with the gain medium in the lasing experiments such, that lasing can be achieved. The size and the shape of the nanoparticles was selected to optimize the excitation of the plasmon resonances. The experimental data are in excellent agreement with the theory. The expected transition from a flat band for small $L$ to the curved TM mode of a square lattice at larger $L$ is clearly visible, and the cross-like TE modes of a square array become visible for $L = 40$. 

The flat band in the single-chain case is intuitively understood as the $x$-direction discrete translational invariance (lattice periodicity) fixing the wavelength of the light while the $y$-direction momentum is not limited since the narrow chain breaks translational invariance in the $y$-direction. In the 2D arrays, there is periodicity also perpendicular to the chains, but with a large unit cell: the TM modes in the replicas of the corresponding small Brillouin zones form the flat bands.  
It can also be formally explained by the light dispersion in a 2D rectangular periodic system
\begin{equation}
\label{e:dispersion_rectangular}
E(\mathbf{k}_\parallel) = \frac{\hbar c_0}{n} \sqrt{\left(k_x + m \frac{2\pi}{a_x}\right)^2 + \left(k_y + m' \frac{2\pi}{a_y}\right)^2},    
\end{equation}
where $\mathbf{k}_\parallel = (k_x, k_y)^\text{T}$, $\hbar$ is the reduced Planck's constant, $c_0$ the speed of light in vacuum, $n$ the refractive index, $a_x$, $a_y$ the periods, and $m$, $m'$ integers. Let us consider the case $E \sim 2\pi\hbar c_0 /(n a_x)$, i.e., diffraction ($m=\pm 1$) in the $x$-direction, and $k_x=0$. The single chain corresponds to $a_y \rightarrow \infty$: there are thus infinitely densely spaced values of $m' 2\pi/a_y$ to match any values of $k_y$ so that the energy remains the same, producing a flat band. For a finite $a_y$, the first diffracted order $m'=\pm 1$ would correspond to quite different energy, but $m'=0$ gives the approximately parabolic dispersion $E=(\hbar c_0/n) \sqrt{(2\pi/a_x)^2 + k_y^2}$ close to the energy $E \sim 2\pi\hbar c_0 /(n a_x)$. In a transition between a single chain and a full 2D lattice, both the flat band and the parabolic TM mode are visible in the experiments and theory of Fig.~\ref{f:dispersion_chains}; both of these are determined by the first diffracted order in the $x$-direction. The cross-type TE mode that appears for $L=40$ corresponds to the first diffracted order in the $y$-direction ($m'=\pm 1$, $m=0$), giving a linear dispersion $E=(\hbar c_0/n)(k_y \pm 2\pi/a_y)$ (in our example $a_x$ and $a_y$ are chosen the same).     

\begin{figure*}[htbp!]
    \centering
    \includegraphics[width=1\textwidth]{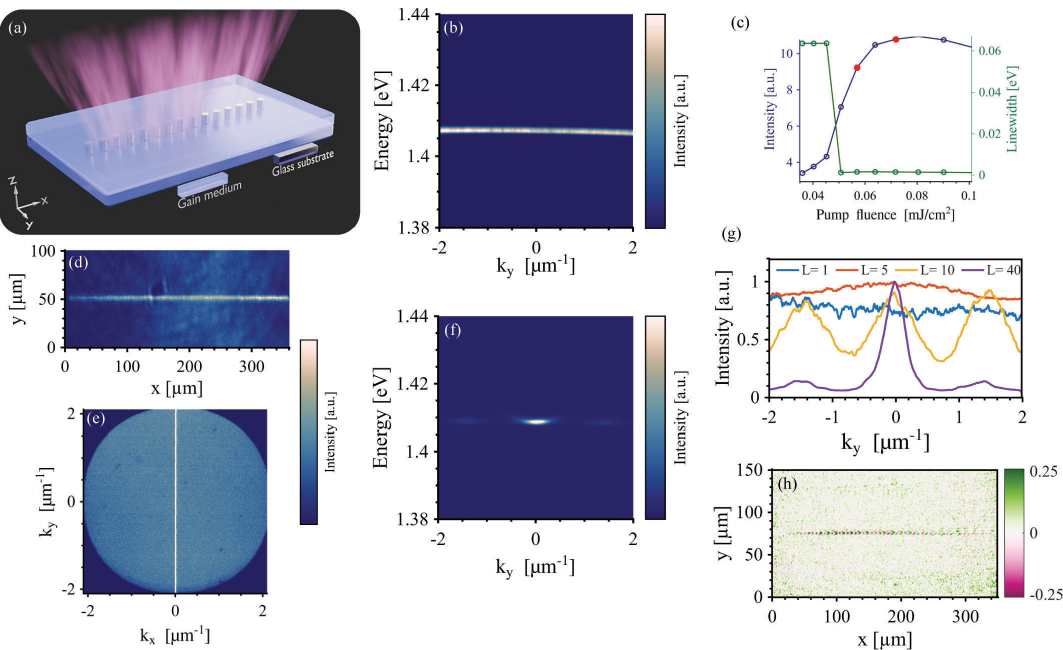}
    \caption{Lasing emission from single chain systems. (a) Scheme of the single chain. (b) Momentum-space-resolved spectrum of single-chain system lasing emission at a pump fluence of 0.05~mJ/cm$^{2}$. (c) Emission intensity and linewidth dependencies on pump fluence for a single-chain system at a pump fluence of 0.07~mJ/cm$^{2}$.  (d) and (e) real and momentum-space emission patterns of single-chain lasing. (f) Momentum-space-resolved spectrum of an $L=40$ chain system lasing emission at a pump fluence of 0.09~mJ/cm$^{2}$. (g) Lasing emission dependence on $k_y$ for systems of different numbers of chains. The pump fluences were as follows: 0.05~mJ/cm$^{2}$ for $L=1$,  0.07~mJ/cm$^{2}$ for $L=5$,  0.1~mJ/cm$^{2}$ for $L=10$, and  0.09~mJ/cm$^{2}$ for $L=40$ (h) Background-free Michelson interference pattern of single-chain lasing emission at a pump fluence of 0.1~mJ/cm$^{2}$.}
    \label{f:single_chain}
\end{figure*}

We study the lasing emission from the flat-band mode by combining the nanoparticle chains with an organic gain medium (IR 140, 10~mM) and pumping this system optically with a 100~fs pulsed laser at a central wavelength of 800~nm (see 
Supplementary Information for details on the sample fabrication and measurements). We first study the single chain system in Fig.~\ref{f:single_chain} to understand lasing within a single building block of the system. The emission spectrum above threshold is shown in Fig.~\ref{f:single_chain}(b), where the emission clearly stems from the flat band mode observed in Fig.~\ref{f:dispersion_chains}(b). With increasing pump fluence, we observe an exponential increase in intensity and a prominent narrowing of the linewidth at threshold [see Fig.~\ref{f:single_chain}(c)], providing a Q-factor of 900. The real space and momentum space emission are shown in Figs.~\ref{f:single_chain}(d) and (e), respectively. From Fig.~\ref{f:single_chain}(e), it is evident that the emission covers the whole range in $k_{y}$ while at the same time being confined tightly around $k_{x} = 0$. We measured the spatial coherence of the single-chain emission via Michelson interferometry; for a description of the method, see the Supplementary Information. 
The emission is coherent as there are clearly visible fringes in the interference pattern [Fig.~\ref{f:single_chain}(h)], hence the system is lasing. We also note that Rekola et al. studied single chains of nanoparticles~\cite{rekola2018one}; however, they focused on the dispersion along the chain (here $k_x$), which does not show a flat band.

In addition to the single chains, we also studied ensembles of chains with different $L$. The emission spectrum of a lattice with $L=40$ is shown in Fig.~\ref{f:single_chain}(f), where most of the emission is now at the $\Gamma$ point, with two side-maxima. We observe that the lasing emission along $k_{y}$ depends strongly on the number $L$ of chains, where the flat band emission breaks down at $L=10$ [Fig.~\ref{f:single_chain}(g)]. The side maxima visible for $L=10$ and $L=40$ are related to the system size in the $y$-direction.

\begin{figure*}[htbp!]
    \centering
    \includegraphics[width=1\linewidth]{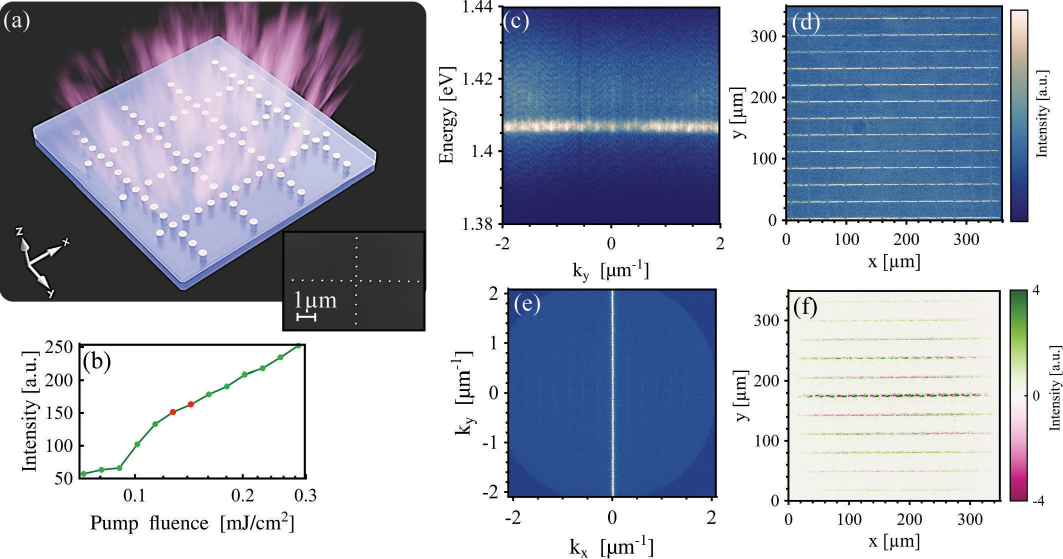}
    \caption{Emission from square 2D chain arrays. (a) SEM image and scheme of the square 2D chain array. (b) Emission intensity dependence on pump fluence. (c) Momentum-space-resolved spectrum of square 2D array emission at a pump fluence of 0.13~mJ/cm$^{2}$. (d) and (e) Real and momentum-space pattern of square 2D chain array emission at a pump fluence of 0.14~mJ/cm$^{2}$. (f) Background-free interference pattern of the square 2D chain array emission, showing fringes along the x-direction but fading features in y. The pump fluence was 0.25~mJ/cm$^{2}$. The arrays consisted of cylindrical gold nanoparticles with a diameter of 110~nm and a height of 50~nm. The period of the array was 580~nm, and the distance between the chains was 40 particles.}
    \label{f:cross}
\end{figure*}

Next, we study combinations of several chains in different configurations to tailor the direction of the flat band and to utilize the gain medium more efficiently than a single chain would. In our first example, chains that are combined to form a square 2D chain array [Fig.~\ref{f:cross}(a)] show under optical pumping a nonlinear increase in emission intensity [Fig.~\ref{f:cross}(b)] with a Q-factor of 200. The emission above threshold stems from the flat band TM mode similar to the single chains, as is evident from the momentum-resolved spectrum and the momentum-space emission pattern in Figs.~\ref{f:cross}(c) and (e), respectively. The flat band mode is formed by the superposition of the SLR modes associated with the diffracted orders ($\pm 1$, $m'$), $m' \in \mathbb{Z}$. The real-space emission pattern shows that the emission originates mainly from the chains in the $x$-direction [see Fig.~\ref{f:cross}(d)]. Although the structure is $x$-$y$ symmetric, one direction is favored due to the pump polarization (along $y$). For more information on the influence of the pump polarization, see Fig. S4. 
The interference pattern of the square 2D chain array [Fig.~\ref{f:cross}(f)] shows clear coherence for the line in the center only. This means that we have coherent emission in part of the system---namely each single line is coherent with itself and hence lasing. However, the system as a whole is not coherent, and emission of different chains is not phase-locked.  

\begin{figure*}[htbp!]
    \centering
    \includegraphics[width=1\linewidth]{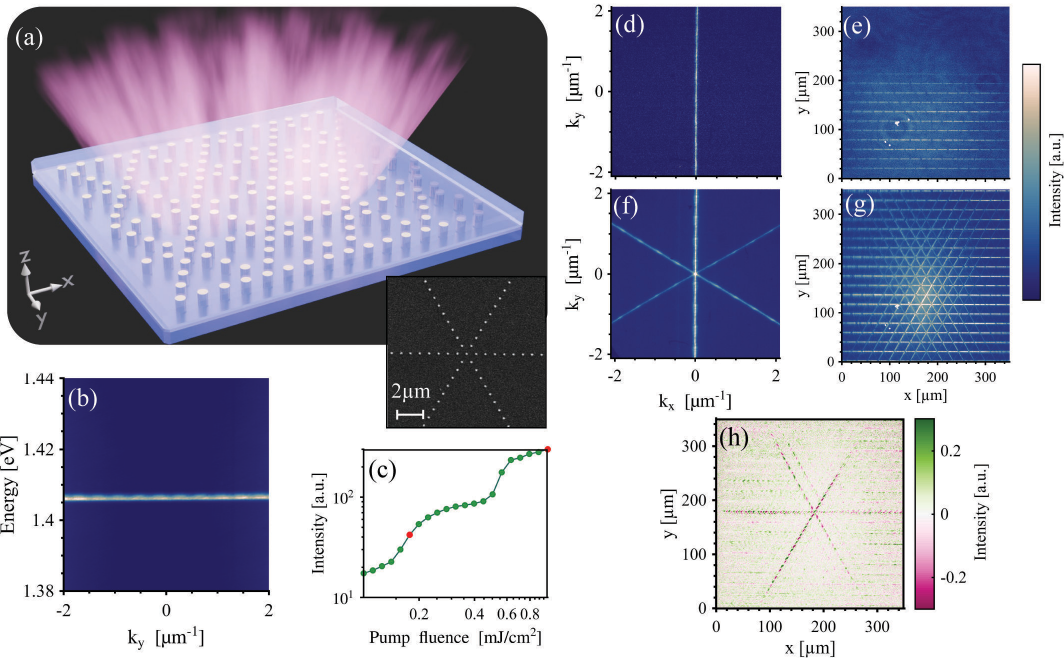}
    \caption{Emission from triangular 2D chain arrays. (a) SEM image and scheme of the triangular 2D chain array. (b) Momentum-resolved spectrum of the triangular array emission at a pump fluence of 0.18~$\mathrm{mJ/cm^2}$. (c) Emission intensity dependence on the pump fluence, the red dots denote pump fluence values used for momentum and real-space pattern collection. (d)--(g) Momentum [(d) and (f)] and real [(e) and (g)] space emission patterns of triangular 2D chain arrays for the fluence values of 0.18~$\mathrm{mJ/cm^2}$ [(d) and (e)] and 1~$\mathrm{mJ/cm^2}$ [(f) and (g)]. (h) Background-free interference pattern of triangular 2D chain array emission under the fluence value of 1 $\mathrm{mJ/cm^2}$. The arrays consisted of cylindrical gold nanoparticles with a diameter of 120~nm and a height of 50~nm. The period of the chains was 580~nm, and the distance between the chains was 33 particles.}
    \label{f:cross_tri}
\end{figure*}

To demonstrate the versatility of the concept, we show in Fig.~\ref{f:cross_tri} the emission of triangular 2D chain arrays, where chains that are oriented along three different angles combine to form an array [see Fig.~\ref{f:cross_tri}(a) for an SEM image and the scheme of the array]. While the momentum-space-resolved spectrum again shows emission from the flat band TM mode [Fig.~\ref{f:cross_tri}(b)], we clearly observe two thresholds in the emission intensity with increasing pump fluence [Fig.~\ref{f:cross_tri}(c)]. The real-space and momentum-space emission patterns are shown for these two regimes in Figs.~\ref{f:cross_tri}(d) and (e) for the lower pump fluence and (f) and (g) for the higher pump fluence regime. In the lower pump fluence, only the chains along the $x$ axis are excited, showing emission only at $k_{x}=0$ and for all $k_y$ values (again, this direction is chosen by the $y$-polarized pump), whereas at the higher pump fluence the diagonal chains are also excited, leading to flat band emission also along other angles than $k_{x}=0$, that is, $k_{y}=\pm \mathrm{tan(}30\mathrm{^{\circ})} k_x$ (these two regimes can also be observed for square 2D chain arrays, see Figs. S4 and S5. 
Again, the central lines are coherent with themselves, meaning that single chains are lasing, whereas the array as a whole shows incoherent emission: this is evident from the interference image shown in Fig.~\ref{f:cross_tri}(h), which has fringes only along positions that correspond to coherence along the chains in three different directions.

In summary, we have experimentally demonstrated that arrays built from single nanoparticle chains are flexible platforms for bright, polarized light generation: the design of the chain arrangement allows realizing different axially symmetric flat band patterns, and applying specific pump energies leads to effective excitation of only certain modes.

We first demonstrated the formation of a flat band in the TM mode of single chains of nanoparticles and experimentally observed lasing in this flat band. We showed that the emission of the single chains is coherent also when combining ensembles of chains into arrays. While the single building blocks are coherent, these parts are not mutually coherent. As such, this is useful as a bright, incoherent light source, while optimization of the Q-factors and chain couplings could be done to make the total emission coherent, if desired. Further, we showed that we can control the angles at which the flat band appears by arranging the chains accordingly. By changing the pump fluence as well as the pump polarization, specific modes of the system can be excited.

Because the flat bands in the chain lattices arise purely from diffraction~\cite{Lehikoinen2026}, their appearance is not tied to a specific experimental platform. Due to the small amount of particles in the chains and 2D chain systems, sufficiently large particles are needed to obtain strong enough modes to experimentally realize flat bands. However, ohmic losses in plasmonic particles prevent lasing in chain systems of particles for a diameter larger than 130~nm.
Conceivably, due to the absence of ohmic losses, higher $Q$-factors could be reached for the flat band modes using dielectric nanoparticles instead of metallic ones; the former also offer an additional design degree of freedom due to their magnetic resonances. Two key properties differentiate the chain lattices from the previously reported flat-band lasing realizations~\cite{baboux2016bosonic, klembt2017polariton, whittaker2018exciton, harder2020exciton, harder2021kagome, scafirimuto2021tunable, Mao2021, Luan2023, Do2025, Eyvazi2025, Cui2025}. First, other realizations of flat-band lasing systems with long-range coupling have relied on using guided modes~\cite{Eyvazi2025, Do2025}, leading to either quasi-flat band systems or flat bands within a finite angular extent. The flat bands reported here extend naturally to all angles. Second, here the appearance of flat bands does not require fine-tuning of the lattice parameters (e.g., to realize nearest-neighbor coupling): instead, the lattice parameters determine the spectral positions of the flat bands predictably and straightforwardly. Our experiments and the theory work of Ref.~\cite{Lehikoinen2026} thus introduce a promising complementary concept for further studies and applications of photonic flat bands. 

\section*{Competing interests}
R. H., J. L., and P. T. declare that they are inventors in a filed PCT patent application (PCT/FI2026/050087) related to the work described in this manuscript. The remaining authors declare no competing interests.

\section*{Acknowledgments}
Funded by the European Union. Views and opinions expressed are however those of the author(s) only and do not necessarily reflect those of the European Union or the European Innovation Council and the SMEs Executive Agency (EISMEA). Neither the European Union nor the granting authority can be held responsible for them (SCOLED, Grant Agreement No. 101098813). The work was also supported by the Jane and Aatos Erkko Foundation and the Technology Industries of Finland Centennial Foundation as part of the Future Makers funding program, by the Research Council of Finland under project number 349313, and by the Research Council of Finland through the Finnish Quantum Flagship project 358877. The work is part of the Research Council of Finland Flagship Programme, Photonics Research and Innovation (PREIN), decision number 346529, Aalto University. This work is part of the Finnish Centre of Excellence in Quantum Materials (QMAT). 

Part of the research was performed at the OtaNano Nanofab cleanroom (Micronova Nanofabrication Centre), supported by Aalto University. 

\section*{Author Contribution}
P.T. initiated and supervised the project. R.H. fabricated the 2D chain lattice samples, performed the lasing experiments on the 2D chain lattices and measured the coherence of the square 2D chain arrays. J.L. performed the empty lattice approximation calculations. S.E. fabricated the single chain samples, performed the experiments on the single chain systems and took the SEM images of all samples. E.A.M. measured the coherence of the single chain and triangular 2D chain arrays. All authors discussed the results and wrote the manuscript together.
\section*{Supporting Information:} Sample fabrication; lasing and transmission measurement setup; Spatial coherence measurement principle; Theory; Influence of pump polarization; $k$-space emission data from square chain lattices; Emission from 2D chain arrays with IR792 dye.

\section*{Data availabilty}
The experimental raw data is available at \url{https://doi.org/10.5281/zenodo.18314743}.

\clearpage

\nocite{dataset}
\clearpage
\bibliography{bibliography}

@article{choi2024nonlocal,
  title={Nonlocal, Flat-Band Meta-Optics for Monolithic, High-Efficiency, Compact Photodetectors},
  author={Choi, Minho and Munley, Christopher and Froch, Johannes E and Chen, Rui and Majumdar, Arka},
  journal={Nano Letters},
  volume={24},
  number={10},
  pages={3150--3156},
  year={2024},
  publisher={ACS Publications},
  url={https://pubs.acs.org/doi/abs/10.1021/acs.nanolett.3c05139}
}

@article{munley2023visible,
  title={Visible wavelength flatband in a gallium phosphide metasurface},
  author={Munley, Christopher and Manna, Arnab and Sharp, David and Choi, Minho and Nguyen, Hao A and Cossairt, Brandi M and Li, Mo and Barnard, Arthur W and Majumdar, Arka},
  journal={ACS Photonics},
  volume={10},
  number={8},
  pages={2456--2460},
  year={2023},
  publisher={ACS Publications},
  url={https://pubs.acs.org/doi/pdf/10.1021/acsphotonics.3c00175}
}

@article{scafirimuto2021tunable,
  title={Tunable exciton--polariton condensation in a two-dimensional {Lieb} lattice at room temperature},
  author={Scafirimuto, Fabio and Urbonas, Darius and Becker, Michael A and Scherf, Ullrich and Mahrt, Rainer F and St{\"o}ferle, Thilo},
  journal={Communications Physics},
  volume={4},
  number={1},
  pages={39},
  year={2021},
  publisher={Nature Publishing Group UK London},
  url={https://www.nature.com/articles/s42005-021-00548-w},
  note = {Accessed: 2026-04-07}
}

@article{harder2021kagome,
  title={Kagome flatbands for coherent exciton-polariton lasing},
  author={Harder, Tristan H and Egorov, Oleg A and Krause, Constantin and Beierlein, Johannes and Gagel, Philipp and Emmerling, Monika and Schneider, Christian and Peschel, Ulf and H{\"o}fling, Sven and Klembt, Sebastian},
  journal={ACS Photonics},
  volume={8},
  number={11},
  pages={3193--3200},
  year={2021},
  publisher={ACS Publications},
  url={https://pubs.acs.org/doi/10.1021/acsphotonics.1c00950}
}

@article{baboux2016bosonic,
  title={Bosonic condensation and disorder-induced localization in a flat band},
  author={Baboux, F and Ge, L and Jacqmin, Thibault and Biondi, M and Galopin, E and Lema{\^\i}tre, A and Le Gratiet, L and Sagnes, Isabelle and Schmidt, S and T{\"u}reci, Hakan E and Amo, A and Bloch, J},
  journal={Physical review letters},
  volume={116},
  number={6},
  pages={066402},
  year={2016},
  publisher={APS},
  url={https://link.aps.org/doi/10.1103/PhysRevLett.116.066402}
}

@article{longhi2019photonic,
  title={Photonic flat-band laser},
  author={Longhi, Stefano},
  journal={Optics letters},
  volume={44},
  number={2},
  pages={287--290},
  year={2019},
  publisher={Optica Publishing Group},
  url={https://opg.optica.org/abstract.cfm?URI=ol-44-2-287},
  note = {Accessed: 2026-04-07}
}

@article{hakala2017lasing,
  title={Lasing in dark and bright modes of a finite-sized plasmonic lattice},
  author={Hakala, TK and Rekola, HT and V{\"a}kev{\"a}inen, AI and Martikainen, J-P and Ne{\v{c}}ada, Marek and Moilanen, AJ and T{\"o}rm{\"a}, P},
  journal={Nature communications},
  volume={8},
  number={1},
  pages={13687},
  year={2017},
  publisher={Nature Publishing Group UK London},
  url={https://www.nature.com/articles/ncomms13687},
  note = {Accessed: 2026-04-07}
}

@article{klembt2017polariton,
  title={Polariton condensation in {S}-and {P}-flatbands in a two-dimensional {Lieb} lattice},
  author={Klembt, Sebastian and Harder, Tristan H and Egorov, Oleg A and Winkler, Karol and Suchomel, Holger and Beierlein, Johannes and Emmerling, Monika and Schneider, Christian and H{\"o}fling, Sven},
  journal={Applied Physics Letters},
  volume={111},
  number={23},
  year={2017},
  publisher={AIP Publishing},
  url={https://pubs.aip.org/apl/article/111/23/231102/34756/Polariton-condensation-in-S-and-P-flatbands-in-a},
  note = {Accessed: 2026-04-07}
}

@article{whittaker2018exciton,
  title={Exciton polaritons in a two-dimensional {Lieb} lattice with spin-orbit coupling},
  author={Whittaker, CE and Cancellieri, Emiliano and Walker, PM and Gulevich, DR and Schomerus, H and Vaitiekus, D and Royall, B and Whittaker, DM and Clarke, E and Iorsh, IV and others},
  journal={Physical review letters},
  volume={120},
  number={9},
  pages={097401},
  year={2018},
  publisher={APS},
  url={https://link.aps.org/doi/10.1103/PhysRevLett.120.097401}
}

@article{harder2020exciton,
  title={Exciton-polaritons in flatland: Controlling flatband properties in a {Lieb} lattice},
  author={Harder, Tristan H and Egorov, Oleg A and Beierlein, Johannes and Gagel, Philipp and Michl, Johannes and Emmerling, Monika and Schneider, Christian and Peschel, Ulf and H{\"o}fling, Sven and Klembt, Sebastian},
  journal={Physical Review B},
  volume={102},
  number={12},
  pages={121302},
  year={2020},
  publisher={APS},
  url={https://link.aps.org/doi/10.1103/PhysRevB.102.121302}
}

@article{leykam2018artificial,
  title={Artificial flat band systems: from lattice models to experiments},
  author={Leykam, Daniel and Andreanov, Alexei and Flach, Sergej},
  journal={Advances in Physics: X},
  volume={3},
  number={1},
  pages={1473052},
  year={2018},
  publisher={Taylor \& Francis},
  url={https://www.tandfonline.com/doi/full/10.1080/23746149.2018.1473052}
}

@article{leykam2018perspective,
  title={Perspective: photonic flatbands},
  author={Leykam, Daniel and Flach, Sergej},
  journal={APL Photonics},
  volume={3},
  number={7},
  year={2018},
  publisher={AIP Publishing},
  url={https://pubs.aip.org/app/article/3/7/070901/123033/Perspective-Photonic-flatbands},
  note = {Accessed: 2026-04-07}
}

@article{Danieli2024,
   author = {Carlo Danieli and Alexei Andreanov and Daniel Leykam and Sergej Flach},
   doi = {10.1515/nanoph-2024-0135},
   issn = {2192-8614},
   issue = {21},
   journal = {Nanophotonics},
   month = {9},
   pages = {3925-3944},
   publisher = {Walter de Gruyter GmbH},
   title = {Flat band fine-tuning and its photonic applications},
   volume = {13},
   year = {2024},
}

@article{Lieb1989,
   author = {Elliott H. Lieb},
   doi = {10.1103/PhysRevLett.62.1201},
   issn = {0031-9007},
   issue = {10},
   journal = {Physical Review Letters},
   month = {3},
   pages = {1201-1204},
   title = {Two theorems on the {H}ubbard model},
   volume = {62},
   url = {https://link.aps.org/doi/10.1103/PhysRevLett.62.1201},
   year = {1989},
}

@article{Eyvazi2025,
   author = {Sioneh Eyvazi and Evgeny A. Mamonov and Rebecca Heilmann and Javier Cuerda and Päivi Törmä},
   doi = {10.1021/acsphotonics.4c02332},
   issn = {2330-4022},
   issue = {3},
   journal = {ACS Photonics},
   month = {3},
   pages = {1570-1578},
   publisher = {American Chemical Society},
   title = {Flat-Band Lasing in Silicon Waveguide-Integrated Metasurfaces},
   volume = {12},
   url = {https://pubs.acs.org/doi/10.1021/acsphotonics.4c02332},
   year = {2025}
}

@article{Mao2021,
   author = {Xin-Rui Mao and Zeng-Kai Shao and Hong-Yi Luan and Shao-Lei Wang and Ren-Min Ma},
   doi = {10.1038/s41565-021-00956-7},
   issn = {1748-3387},
   issue = {10},
   journal = {Nature Nanotechnology},
   month = {10},
   pages = {1099-1105},
   pmid = {34400821},
   publisher = {Nature Research},
   title = {Magic-angle lasers in nanostructured moiré superlattice},
   volume = {16},
   url = {https://www.nature.com/articles/s41565-021-00956-7},
   year = {2021},
  note = {Accessed: 2026-04-07}
}

@article{Luan2023,
   author = {Hong-Yi Luan and Yun-Hao Ouyang and Zi-Wei Zhao and Wen-Zhi Mao and Ren-Min Ma},
   doi = {10.1038/s41586-023-06789-9},
   issn = {0028-0836},
   issue = {7991},
   journal = {Nature},
   month = {12},
   pages = {282-288},
   pmid = {38092911},
   publisher = {Nature Research},
   title = {Reconfigurable moiré nanolaser arrays with phase synchronization},
   volume = {624},
   url = {https://www.nature.com/articles/s41586-023-06789-9},
   year = {2023},
  note = {Accessed: 2026-04-07}
}

@article{Ning2023,
   author = {Tingyin Ning and Lina Zhao and Yanyan Huo and Yangjian Cai and Yingying Ren},
   doi = {10.1515/nanoph-2023-0124},
   issn = {2192-8614},
   issue = {21},
   journal = {Nanophotonics},
   month = {10},
   pages = {4009-4016},
   publisher = {Walter de Gruyter GmbH},
   title = {Giant enhancement of second harmonic generation from monolayer {2D} materials placed on photonic moiré superlattice},
   volume = {12},
   url = {https://www.degruyter.com/document/doi/10.1515/nanoph-2023-0124/html},
   year = {2023}
}

@article{Do2025,
   author = {Thi Thu Ha Do and Zhiyi Yuan and Emek G. Durmusoglu and Hadi K. Shamkhi and Vytautas Valuckas and Chunyu Zhao and Arseniy I. Kuznetsov and Hilmi Volkan Demir and Cuong Dang and Hai Son Nguyen and Son Tung Ha},
   doi = {10.1021/acsnano.5c01972},
   issn = {1936-0851},
   issue = {20},
   journal = {ACS Nano},
   month = {5},
   pages = {19287-19296},
   title = {Room-Temperature Lasing at Flatband Bound States in the Continuum},
   volume = {19},
   url = {https://pubs.acs.org/doi/10.1021/acsnano.5c01972},
   year = {2025}
}

@misc{Lehikoinen2026,
  author = {Joel Lehikoinen and Rebecca Heilmann and Aron J. J. Dahlberg and Eero Härmä and Malek Mahmoudi and Arpan Dutta and Konstantinos S. Daskalakis and Päivi Törmä},
  title  = {Flat Bands from Diffraction in Periodic Systems},
  note   = {Accessed: 2026-04-07. Submitted: 2026-02-25 to arXiv:2602.21830 [physics.optics]},
  year = {2026},
  URL={https://arxiv.org/abs/2602.21830}
}

@article{Cui2025,
   author = {Jieyuan Cui and Song Han and Bofeng Zhu and Chongwu Wang and Yunda Chua and Qian Wang and Lianhe Li and Alexander Giles Davies and Edmund Harold Linfield and Qi Jie Wang},
   doi = {10.1038/s41566-025-01665-6},
   issn = {1749-4885},
   issue = {6},
   journal = {Nature Photonics},
   month = {6},
   pages = {643-649},
   publisher = {Nature Research},
   title = {Ultracompact multibound-state-assisted flat-band lasers},
   volume = {19},
   url = {https://www.nature.com/articles/s41566-025-01665-6},
   year = {2025},
  note = {Accessed: 2026-04-07}
}

@article{Sun2025,
   author = {Kaili Sun and Yangjian Cai and Yuri Kivshar and Zhanghua Han},
   issn = {2577-5421},
   issue = {05},
   journal = {Advanced Photonics},
   month = {10},
   title = {Flatband high-{Q} metasurfaces inspired by coupled-resonator optical waveguides},
   volume = {7},
   url = {https://www.spiedigitallibrary.org/journals/advanced-photonics/volume-7/issue-05/056008/Flatband-high-Q-metasurfaces-inspired-by-coupled-resonator-optical-waveguides/10.1117/1.AP.7.5.056008.full},
   year = {2025},
  note = {Accessed: 2026-04-07}
}

@article{rekola2018one,
  title={One-dimensional plasmonic nanoparticle chain lasers},
  author={Rekola, Heikki T and Hakala, Tommi K and T{\"o}rm{\"a}, Päivi},
  journal={ACS photonics},
  volume={5},
  number={5},
  pages={1822--1826},
  year={2018},
  publisher={ACS Publications},
  url={https://pubs.acs.org/doi/10.1021/acsphotonics.8b00001}
}

@article{Bergholtz2013,
    author = {Bergholtz, Emil J. and Liu, Zhao},
    title = {Topological flat band models and fractional {C}hern insulators},
    journal = {International Journal of Modern Physics B},
    volume = {27},
    number = {24},
    pages = {1330017},
    year = {2013},
    doi = {10.1142/S021797921330017X},
    URL = {https://doi.org/10.1142/S021797921330017X}
}

@article{Torma2022,
   author = {Päivi Törmä and Sebastiano Peotta and Bogdan A. Bernevig},
   doi = {10.1038/s42254-022-00466-y},
   issn = {2522-5820},
   issue = {8},
   journal = {Nature Reviews Physics},
   month = {6},
   pages = {528-542},
   publisher = {Springer Nature},
   title = {Superconductivity, superfluidity and quantum geometry in twisted multilayer systems},
   volume = {4},
   url = {https://www.nature.com/articles/s42254-022-00466-y},
   year = {2022}
}

@article{Baba2008,
   author = {Toshihiko Baba},
   doi = {10.1038/nphoton.2008.146},
   issn = {1749-4885},
   issue = {8},
   journal = {Nature Photonics},
   month = {8},
   pages = {465-473},
   title = {Slow light in photonic crystals},
   volume = {2},
   url = {https://www.nature.com/articles/nphoton.2008.146},
   year = {2008},
  note = {Accessed: 2026-04-07}
}

@article{Kravets2018,
   author = {V. G. Kravets and A. V. Kabashin and W. L. Barnes and A. N. Grigorenko},
   doi = {10.1021/acs.chemrev.8b00243},
   issn = {0009-2665},
   issue = {12},
   journal = {Chemical Reviews},
   month = {6},
   pages = {5912-5951},
   publisher = {American Chemical Society},
   title = {Plasmonic Surface Lattice Resonances: A Review of Properties and Applications},
   volume = {118},
   url = {https://pubs.acs.org/doi/10.1021/acs.chemrev.8b00243},
   year = {2018}
}

@Article{Suh2012,
    author={Suh, Jae Yong and Kim, Chul Hoon and Zhou, Wei and Huntington, Mark D. and Co, Dick T. and Wasielewski, Michael R. and Odom, Teri W.},
    title={Plasmonic Bowtie Nanolaser Arrays},
    journal={Nano Letters},
    year={2012},
    month={Nov},
    day={14},
    publisher={American Chemical Society},
    volume={12},
    number={11},
    pages={5769-5774},
    issn={1530-6984},
    doi={10.1021/nl303086r},
    url={https://doi.org/10.1021/nl303086r}
}

@Article{Zhou2013,
    author={Zhou, Wei and Dridi, Montacer and Suh, Jae Yong and Kim, Chul Hoon and Co, Dick T. and Wasielewski, Michael R. and Schatz, George C. and Odom, Teri W.},
    title={Lasing action in strongly coupled plasmonic nanocavity arrays},
    journal={Nature Nanotechnology},
    year={2013},
    month={Jul},
    day={01},
    volume={8},
    number={7},
    pages={506-511},
    issn={1748-3395},
    doi={10.1038/nnano.2013.99},
    url={https://doi.org/10.1038/nnano.2013.99}
}

@article{wang2018rich,
  title={The rich photonic world of plasmonic nanoparticle arrays},
  author={Wang, Weijia and Ramezani, Mohammad and V{\"a}kev{\"a}inen, Aaro I and T{\"o}rm{\"a}, P{\"a}ivi and Rivas, Jaime G{\'o}mez and Odom, Teri W},
  journal={Materials today},
  volume={21},
  number={3},
  pages={303--314},
  year={2018},
  publisher={Elsevier},
  url={https://linkinghub.elsevier.com/retrieve/pii/S1369702117306727},
  note = {Accessed: 2026-04-07}
}

@article{freire2025plasmonic,
  title={Plasmonic lattice lasers},
  author={Freire-Fern{\'a}ndez, Francisco and Park, Sang-Min and Tan, Max JH and Odom, Teri W},
  journal={Nature Reviews Materials},
  pages={1--13},
  year={2025},
  publisher={Nature Publishing Group UK London},
  url={https://www.nature.com/articles/s41578-025-00803-4},
  note = {Accessed: 2026-04-07}
}

@misc{dataset,
  author       = {Rebecca Heilmann and Joel Lehikoinen and Sioneh Eyvazi and Evgeny A. Mamonov and Päivi Törmä},
  title        = {Data for: Chains of nanoparticles for flat band emission and lasing},
  year         = {2026},
  publisher    = {Zenodo},
  doi          = {10.5281/zenodo.18314743},
  url          = {https://doi.org/10.5281/zenodo.18314743}
}

@article{Hoang2024,
   author = {Thanh Xuan Hoang and Daniel Leykam and Yuri Kivshar},
   doi = {10.1103/PhysRevLett.132.043803},
   issn = {0031-9007},
   issue = {4},
   journal = {Physical Review Letters},
   month = {1},
   pages = {043803},
   publisher = {American Physical Society},
   title = {Photonic Flatband Resonances in Multiple Light Scattering},
   volume = {132},
   url = {https://link.aps.org/doi/10.1103/PhysRevLett.132.043803},
   year = {2024}
}

\newpage


\section*{Supplementary material}

coherence of the single chain and triangular 2D chain arrays. All authors discussed the results and wrote the manuscript together.

## Methods

### Sample fabrication

The nanoparticle arrays were fabricated with electron beam lithography on borosilicate substrates ($n = 1.52$). A 2 nm layer of titanium was used for better adhesion of the gold nanoparticles to the substrates. The height of the nanoparticles was 50 nm, the particle diameter, period, and length of the single chains were 120 nm, 580 nm, and 300 µm. The particle diameter, period, and size of the 2D chain arrays were 110 nm, 580 nm, and 300 µm for the square 2D chain arrays; the distance between the chains was 40 particles. The particle diameter, array period, and size for the triangular 2D chain arrays were 120 nm, 580 nm, and 300 µm, where the distance of the chains was 33 particles. For the dispersion measurements, the samples were immersed in index-matching oil and sealed with a cover slip. For lasing experiments, a dye solution of IR140 (concentration of 10 mM) or IR792 (concentration of 30 mM) in a 2:1 mixture of benzyl alcohol and dimethyl sulfoxide was prepared. The solvent mixture was chosen such that the refractive index of the dye solution matches that of the substrate. The arrays were immersed in the dye solution by injecting the solution into a cavity of a press-to-seal silicone ring (0.8 mm thickness) pressed between the substrate and a second borosilicate slide.

### Lasing and transmission measurement setup

The schematic of the experimental setup used for transmission and lasing experiments is shown in Fig. S1. In transmission measurements, the sample is illuminated with a white light source. The signal from the array is imaged with an NA 0.3 objective, whose back focal plane is imaged onto a spectrometer slit. The back focal plane contains the information about angular distribution of the emission and hence each position of the slit corresponds to a specific angle $\theta_y$, related to the in-plane wavevector by $k_y = k_0\sin(\theta_y)$, where $k_0 = 2\pi/\lambda_0$, with the free space wavelength $\lambda_0$. The grating inside the spectrometer disperses the light onto a 2D CCD sensor, which then resolves the wavelength along one axis and the wavevector along the other. In addition, real and momentum space images were taken with two CMOS cameras. In the lasing measurements, the sample is optically pumped with ultrafast laser pulses (800 nm central wavelength, 1 kHz repetition rate) through the objective. A long pass filter (cut-off wavelength 850 nm) was used in the detection path to filter out the pump signal. The pump polarization is controlled by a combination of a half-wave plate and a linear polarizer, whereas a neutral density wheel is used to control the pump fluence.

### Spatial coherence measurements

The spatial coherence was measured by routing the emission from the sample with a mirror to a Michelson setup, see Fig. S2.Here, the light is split into two paths, with the first ending on a retroreflector that flips

the real-space emission image along the vertical and horizontal axes. Along the second path, the image is guided to a mirror on a movable stage. Both images, i.e., the flipped and unflipped image of the real space emission, are focused onto a CMOS camera, where the interference image is collected. Mirror M is moved until fringes are most visible in the interference image, indicating the point of zero time delay. Since the retroreflector flips the image of the real space image with respect to the horizontal and vertical axes, we probe coherence of the real space emission between points $(x, y)$ and $(-x, -y)$, as indicated by the crosses in Fig. S3 (a). In total, three measurements are taken: the image of the retroreflector only (blocking the path to Mirror M [Fig. S3 (b)] and the image of the mirror only (blocking the path to the retroreflector) [Fig. S3 (c)]. The sum of these measurements constitutes background emission [Fig. S3 (d)], i.e., no interference effects since one of the beams was blocked in each measurement. The third measurement taken is the interference measurement, where both paths are open Fig. S3 (e). While the fringes are already faintly visible in the raw data of interference, they become clearer after background removal [Fig. S3 (f)]. The visibility of fringes at different locations of this image proves coherence between two specific locations in the original emission, see Figures S3 (a) and (f) for an example.

## Theory

The empty lattice approximation consists of imposing the periodicity of the lattice on the free space dispersion of light and of weighting the contribution of each mode with its structure factor [32]. The structure factor $S(\mathbf{k})$ of the mode characterized by the wavevector $\mathbf{k}$ is defined as

$$S(\mathbf{k}) = \frac{1}{N^2} \left| \sum_{n=0}^{N-1} e^{i\mathbf{k}\cdot\mathbf{r}} \right|^2, \tag{2}$$

where $N$ is the number of lattice sites and $\mathbf{r}_n$ denotes the position of the $n$th lattice site. The choice of the normalization factor $N^{-2}$ ensures that the structure factor takes values between zero and one. The dispersion of light in a homogeneous, linear, isotropic optical background for in-plane modes is

$$E = \frac{\hbar c_0}{n} \sqrt{k_x^2 + k_y^2}, \tag{3}$$

where $\hbar$ is the reduced Planck constant, $c_0$ the speed of light in vacuum, and $n$ the refractive index of the background medium. In the $(k_x, k_y, E)$ space, this dispersion describes the surface of a light cone emanating from the point $(k_x, k_y, E = 0)$. Thus, modes that can scatter light to a particular point $(k_x, k_y, E)$ lie on a ring centered at $(k_x, k_y)$ with the radius determined by $E$. The total scattering intensity is thus proportional to the weighted sum of these modes

$$I(\mathbf{k}_\parallel, E) \propto \int S(\mathbf{k}') \delta(|\mathbf{k}' - \mathbf{k}_\parallel| - \kappa) d\mathbf{k}', \tag{4}$$

where the integration is over the $k_x$-$k_y$ plane, $\mathbf{k}_{\parallel} = (k_x, k_y)^{\mathrm{T}}$, and $\kappa = \frac{n}{\hbar c_0} E$ is the wavevector magnitude corresponding to the energy $E$. Numerically, the above integral is computed by discretizing the $k$ space and summing $S(\mathbf{k})$ over the points for which $||\mathbf{k}' - \mathbf{k}_{\parallel}| - \kappa| \leq \frac{\Delta E}{2}$, where $\Delta E$ is the energy discretization parameter.

If the lattice is taken to be infinite and periodic, the structure factor can take non-zero values only at the sites of the reciprocal lattice $\mathbf{k} = q_1 \mathbf{b_1} + q_2 \mathbf{b_2}$, $q_1, q_2, \in \mathbb{Z}$, where $\mathbf{b}_{1,2}$ are the reciprocal lattice vectors [33]. The structure factor is then only computed at the sites of the reciprocal lattice. For periodic and infinite lattices, we may forgo computing the integral of Eq. (4) and instead visualize the band structure of the lattice by plotting the dispersion given by Eq. (1) for each mode $(m 2\pi/a_x, m' 2\pi/a_y)$. In addition, we indicate the polarization of the mode $(k_x, k_y)$ by its TE-polarization fraction [24]

$$p_{\mathrm{TE}} \equiv \begin{cases} 1, \text{ if } k_x = k_y = 0, \\ \frac{|k_y|}{|k_y|+|k_x|}, \text{ otherwise.} \end{cases} \quad (5)$$

This definition assumes the plane of incidence to be the $yz$ plane; if the $xz$ plane is taken to be the plane of incidence, the $x$ and $y$ components are swapped. For dipolar scatterers, such as small plasmonic nanoparticles, diffraction in (out) of the plane of incidence corresponds to linearly TE (TM) polarized light.

All empty lattice calculations have been carried out using an in-house-developed code.

# References

[1] Leykam, D., Andreanov, A. & Flach, S. Artificial flat band systems: from lattice models to experiments. *Advances in Physics: X* **3**, 1473052 (2018). URL https://www.tandfonline.com/doi/full/10.1080/23746149.2018.1473052.

[2] Leykam, D. & Flach, S. Perspective: photonic flatbands. *Apl Photonics* **3** (2018). URL https://pubs.aip.org/app/article/3/7/070901/123033/Perspective-Photonic-flatbands.

[3] Danieli, C., Andreanov, A., Leykam, D. & Flach, S. Flat band fine-tuning and its photonic applications. *Nanophotonics* **13**, 3925–3944 (2024).

[4] Bergholtz, E. J. & Liu, Z. Topological flat band models and fractional Chern insulators. *International Journal of Modern Physics B* **27**, 1330017 (2013). URL https://doi.org/10.1142/S021797921330017X.

[5] Törmä, P., Peotta, S. & Bernevig, B. A. Superconductivity, superfluidity and quantum geometry in twisted multilayer systems. *Nature Reviews Physics* **4**, 528–542 (2022). URL https://www.nature.com/articles/s42254-022-00466-y.

[6] Lieb, E. H. Two theorems on the Hubbard model. *Physical Review Letters* **62**, 1201–1204 (1989). URL https://link.aps.org/doi/10.1103/PhysRevLett.62.1201.

[7] Ning, T., Zhao, L., Huo, Y., Cai, Y. & Ren, Y. Giant enhancement of second harmonic generation from monolayer 2D materials placed on photonic moiré superlattice. *Nanophotonics* **12**, 4009–4016 (2023). URL https://www.degruyter.com/document/doi/10.1515/nanoph-2023-0124/html.

[8] Sun, K., Cai, Y., Kivshar, Y. & Han, Z. Flatband high-Q metasurfaces inspired by coupled-resonator optical waveguides. *Advanced Photonics* **7** (2025). URL https://www.spiedigitallibrary.org/journals/advanced-photonics/volume-7/issue-05/056008/Flatband-high-Q-metasurfaces-inspired-by-coupled-resonator-optical-waveguides/10.1117/1.AP.7.5.056008.full.

[9] Baba, T. Slow light in photonic crystals. *Nature Photonics* **2**, 465–473 (2008). URL https://www.nature.com/articles/nphoton.2008.146.

[10] Munley, C., Manna, A., Sharp, D., Choi, M., Nguyen, H. A., Cossairt, B. M., Li, M., Barnard, A. W. & Majumdar, A. Visible wavelength flatband in a gallium phosphide metasurface. *ACS Photonics* **10**, 2456–2460 (2023). URL https://pubs.acs.org/doi/pdf/10.1021/acsphotonics.3c00175.

[11] Choi, M., Munley, C., Froch, J. E., Chen, R. & Majumdar, A. Nonlocal, flat-band meta-optics for monolithic, high-efficiency, compact photodetectors. *Nano Letters* **24**, 3150–3156 (2024). URL https://pubs.acs.org/doi/abs/10.1021/acs.nanolett.3c05139.

[12] Baboux, F., Ge, L., Jacqmin, T., Biondi, M., Galopin, E., Lemaître, A., Le Gratiet, L., Sagnes, I., Schmidt, S., Türeci, H. E., Amo, A. & Bloch, J. Bosonic condensation and disorder-induced localization in a flat band. *Physical review letters* **116**, 066402 (2016). URL https://link.aps.org/doi/10.1103/PhysRevLett.116.066402.

[13] Klembt, S., Harder, T. H., Egorov, O. A., Winkler, K., Suchomel, H., Beierlein, J., Emmerling, M., Schneider, C. & Höfling, S. Polariton condensation in S-and P-flatbands in a two-dimensional Lieb lattice. *Applied Physics Letters* **111** (2017). URL https://pubs.aip.org/apl/article/111/23/231102/34756/Polariton-condensation-in-S-and-P-flatbands-in-a.

[14] Whittaker, C., Cancellieri, E., Walker, P., Gulevich, D., Schomerus, H., Vaitiekus, D., Royall, B., Whittaker, D., Clarke, E., Iorsh, I. *et al.* Exciton polaritons in a two-dimensional Lieb lattice with spin-orbit coupling. *Physical review letters* **120**, 097401 (2018). URL https://link.aps.org/doi/10.1103/PhysRevLett.120.097401.

[15] Harder, T. H., Egorov, O. A., Beierlein, J., Gagel, P., Michl, J., Emmerling, M., Schneider, C., Peschel, U., Höfling, S. & Klembt, S. Exciton-polaritons in flatland: Controlling flatband properties in a Lieb lattice. *Physical Review B* **102**, 121302 (2020). URL https://link.aps.org/doi/10.1103/PhysRevB.102.121302.

[16] Harder, T. H., Egorov, O. A., Krause, C., Beierlein, J., Gagel, P., Emmerling, M., Schneider, C., Peschel, U., Höfling, S. & Klembt, S. Kagome flatbands for coherent exciton-polariton lasing. *ACS Photonics* **8**, 3193–3200 (2021). URL https://pubs.acs.org/doi/10.1021/acsphotonics.1c00950.

[17] Scafirimuto, F., Urbonas, D., Becker, M. A., Scherf, U., Mahrt, R. F. & Stöferle, T. Tunable exciton–polariton condensation in a two-dimensional Lieb lattice at room temperature. *Communications Physics* **4**, 39 (2021). URL https://www.nature.com/articles/s42005-021-00548-w.

[18] Longhi, S. Photonic flat-band laser. *Optics letters* **44**, 287–290 (2019). URL https://opg.optica.org/abstract.cfm?URI=ol-44-2-287.

[19] Mao, X.-R., Shao, Z.-K., Luan, H.-Y., Wang, S.-L. & Ma, R.-M. Magic-angle lasers in nanostructured moiré superlattice. *Nature Nanotechnology* **16**, 1099–1105 (2021). URL https://www.nature.com/articles/s41565-021-00956-7.

[20] Luan, H.-Y., Ouyang, Y.-H., Zhao, Z.-W., Mao, W.-Z. & Ma, R.-M. Reconfigurable moiré nanolaser arrays with phase synchronization. *Nature* **624**, 282–288 (2023). URL https://www.nature.com/articles/s41586-023-06789-9.

[21] Eyvazi, S., Mamonov, E. A., Heilmann, R., Cuerda, J. & Törmä, P. Flat-band lasing in silicon waveguide-integrated metasurfaces. *ACS Photonics* **12**, 1570–1578 (2025). URL https://pubs.acs.org/

doi/10.1021/acsphotonics.4c02332.

[22] Do, T. T. H., Yuan, Z., Durmusoglu, E. G., Shamkhi, H. K., Valuckas, V., Zhao, C., Kuznetsov, A. I., Demir, H. V., Dang, C., Nguyen, H. S. & Ha, S. T. Room-temperature lasing at flatband bound states in the continuum. *ACS Nano* **19**, 19287–19296 (2025). URL https://pubs.acs.org/doi/10.1021/acsnano.5c01972.

[23] Cui, J., Han, S., Zhu, B., Wang, C., Chua, Y., Wang, Q., Li, L., Davies, A. G., Linfield, E. H. & Wang, Q. J. Ultracompact multibound-state-assisted flat-band lasers. *Nature Photonics* **19**, 643–649 (2025). URL https://www.nature.com/articles/s41566-025-01665-6.

[24] Lehikoinen, J., Heilmann, R., Dahlberg, A. J. J., Härmä, E., Mahmoudi, M., Dutta, A., Daskalakis, K. S. & Törmä, P. Flat bands from diffraction in periodic systems (2026). Submitted.

[25] Kravets, V. G., Kabashin, A. V., Barnes, W. L. & Grigorenko, A. N. Plasmonic surface lattice resonances: A review of properties and applications. *Chemical Reviews* **118**, 5912–5951 (2018). URL https://pubs.acs.org/doi/10.1021/acs.chemrev.8b00243.

[26] Wang, W., Ramezani, M., Väkeväinen, A. I., Törmä, P., Rivas, J. G. & Odom, T. W. The rich photonic world of plasmonic nanoparticle arrays. *Materials today* **21**, 303–314 (2018). URL https://linkinghub.elsevier.com/retrieve/pii/S1369702117306727.

[27] Suh, J. Y., Kim, C. H., Zhou, W., Huntington, M. D., Co, D. T., Wasielewski, M. R. & Odom, T. W. Plasmonic bowtie nanolaser arrays. *Nano Letters* **12**, 5769–5774 (2012). URL https://doi.org/10.1021/nl303086r.

[28] Zhou, W., Dridi, M., Suh, J. Y., Kim, C. H., Co, D. T., Wasielewski, M. R., Schatz, G. C. & Odom, T. W. Lasing action in strongly coupled plasmonic nanocavity arrays. *Nature Nanotechnology* **8**, 506–511 (2013). URL https://doi.org/10.1038/nnano.2013.99.

[29] Hakala, T., Rekola, H., Väkeväinen, A., Martikainen, J.-P., Nečada, M., Moilanen, A. & Törmä, P. Lasing in dark and bright modes of a finite-sized plasmonic lattice. *Nature communications* **8**, 13687 (2017). URL https://www.nature.com/articles/ncomms13687.

[30] Freire-Fernández, F., Park, S.-M., Tan, M. J. & Odom, T. W. Plasmonic lattice lasers. *Nature Reviews Materials* 1–13 (2025). URL https://www.nature.com/articles/s41578-025-00803-4.

[31] Rekola, H. T., Hakala, T. K. & Törmä, P. One-dimensional plasmonic nanoparticle chain lasers. *ACS photonics* **5**, 1822–1826 (2018). URL https://pubs.acs.org/doi/10.1021/acsphotonics.8b00001.

[32] Guo, R., Hakala, T. K. & Törmä, P. Geometry dependence of surface lattice resonances in plasmonic nanoparticle arrays. *Physical Review B* **95**, 155423 (2017). URL http://link.aps.org/doi/10.1103/

PhysRevB.95.155423.

[33] Egami, T. & Billinge, S. *Underneath the Bragg Peaks* 1st edn, Vol. 6 (Elsevier, 2003). URL https://linkinghub.elsevier.com/retrieve/pii/S1369702103006357.

[34] Heilmann, R., Lehikoinen, J., Eyvazi, S., Mamonov, E. A. & Törmä, P. Data for: Chains of nanoparticles for flat band emission and lasing (2026). URL https://doi.org/10.5281/zenodo.18314743.






\end{document}